\documentclass[manuscript]{acmart}
\usepackage{graphicx} 

\usepackage{ulem}

\title{The Impact of Revealing Large Language Model Stochasticity on Trust, Reliability, and Anthropomorphization}

\author{Chelse Swoopes}
\email{cswoopes@g.harvard.edu}
\affiliation{%
  \institution{Harvard University}
  \country{USA}
}

\author{Tyler Holloway}
\email{tylerholloway@g.harvard.edu}
\affiliation{%
  \institution{Harvard University}
  \country{USA}
}

\author{Elena L. Glassman}
\email{glassman@seas.harvard.edu}
\affiliation{%
  \institution{Harvard University}
  \country{USA}
}

\thanks{TREW Workshop at CHI ’24, May 11, 2024, Honolulu, HI, USA}

\begin{abstract}

Interfaces for interacting with large language models (LLMs) are often designed to mimic human conversations, typically presenting a single response to user queries. This design choice can obscure the probabilistic and predictive nature of these models, potentially fostering undue trust and over-anthropomorphization of the underlying model. In this paper, we investigate (i) the effect of displaying multiple responses simultaneously as a countermeasure to these issues, and (ii) how a cognitive support mechanism--highlighting structural and semantic similarities across responses--helps users deal with the increased cognitive load of that intervention. We conducted a within-subjects study in which participants inspected responses generated by an LLM under three conditions: one response, ten responses with cognitive support, and ten responses without cognitive support. Participants then answered questions about workload, trust and reliance, and anthropomorphization. We conclude by reporting the results of these studies and discussing future work and design opportunities for future LLM interfaces.
\end{abstract}

\begin{document}

\maketitle

\section{Introduction}
Large language models (LLMs) are increasingly being harnessed to power applications and interfaces across various domains. One popular mode of interaction with these LLMs is as a chatbot, i.e., a textual interface with built-in turn-taking that mimics human conversation---providing a single response per interaction. While this design choice may enhance user experience by providing engaging interactivity, it may also invite anthropomorphism more than other interface possibilities. In fact, recently, the CEO of a company with a large LLM development operation, i.e., Meta, was quoted by a major news outlet as suggesting that LLM-powered assistants could serve as mentors, coaches, and cheerleaders---which are roles typically played by fellow humans~\cite{vice_zuckerberg}.

This anthropomorphism can in turn impact users' trust in a system, such as increasing their ``trust resilience, [i.e., having] a higher resistance to breakdowns in trust''~\cite{de2016almost} that may be more than is warranted. 
Previous work has shown that AI systems can be granted undue trust by their users, who can then fail to reason analytically about the AI's recommendations and instead overrely on its answers~\cite{tothinkortrust}. The issue of undue trust and overreliance on LLMs is increasingly important as these models may become integrated into decision-making processes across a range of sectors.


LLMs can generate mulitple varying plausible responses to the same prompt because they sample from a probability distribution over the set of possible words or phrases that they have learned during training. But when users are shown a single, definitive-looking response, they may develop an incorrect mental model of the system, assuming that the LLM operates with a high degree of certainty and has a single correct answer to each query. We hypothesize that users' mental models of the system's behavior when they do not have access to that variability may result in anthropomorphization and poor trust calibration---which have been found to be strongly correlated~\cite{jensen2021trust}. In other words, current interface designs that mimic human conversation may inadvertently encourage users to view these models as knowledgeable experts rather than probabilistic text prediction tools with limitations. 


Existing research has yet to fully explore how different interface designs can mitigate these risks, leaving a significant gap in our understanding of how to construct LLM interfaces that users can more appropriately rely (and not over-rely) on. 
We introduce a design intervention that abandons the typical design choice of one response at a time, as if another human were typing back: showing multiple LLM-generated responses to the same question simultaneously. We assess the impact of this intervention on measures of both user anthropomorphization and trust in the LLM. Given the increase in the amount of text that this design intervention creates, we introduce a second design intervention to mitigate the potential increased cognitive load of the first intervention---cognitive support within the rendering of these responses designed to help the user consume the multiple LLM-generated answers to the same question more effectively. Specifically, we investigate (i) the effects of simultaneously displaying multiple responses generated from a single model to a single prompt, and (ii) how a cognitive support mechanism, highlighting structural and semantic similarities across multiple model responses first demonstrated in~\cite{gero2024supporting}, can help users manage the cognitive load of this intervention. 

To evaluate these questions, we conducted a within-subjects user study in which participants were tasked with inspecting responses generated by a single LLM under three conditions: one response, ten responses without cognitive support, and ten responses with cognitive support. Participants then answered questions about workload using the NASA-TLX scale, trust and reliance using the Human-Computer Trust Scale (HCT), and anthropomorphization using the Godspeed Questionnaire. While we did not find statistically significant quantitative results, we did find compelling themes from the semi-structured interviews that followed the controlled study portion of each session. These participant reflections offer insights for future research on the design of LLM interfaces.

\section{Related Work}
   \subsection{Anthropomorphization}
Anthropomorphism is the inclination to ascribe human-like traits to non-human entities. Previous studies have found that certain system characteristics can cause users to anthropomorphize systems, such as voice ~\cite{nass1993anthropomorphism}, language use ~\cite{laban2021perceptions, schanke2021estimating}, physical appearance that mimic human features ~\cite{riek2009anthropomorphism, bi2023create}, and social cues such demonstrating active listening via echo utterance ~\cite{lee2020perceiving, feine2019taxonomy}.

Anthropomorphization is increasingly observed in users of commercial language model-powered systems and interfaces, such as ChatGPT ~\cite{chatgpt}, and Perplexity ~\cite{perplexity-ai}, not only in academia but also in mainstream media ~\cite{nyt-ai-human-reasoning, nyt-gpt-reminds}. These commercial language model-powered interfaces typically provide a single response to a given prompt, with the exception of Google Bard, which can return three responses simultaneously ~\cite{bard}. In this work, we collect preliminary evidence on how generating multiple simultaneous responses—which (1) only a machine could generate simultaneously, revealing (2) the otherwise invisible variability in LLM-generated answers—affect users’ tendency to anthropomorphize and/or trust these systems.

While there have been efforts to both increase and reduce the anthropomorphization of non-language model powered AI systems and interfaces ~\cite{eyssel2012if, sheehan2020customer}, research on inventions specifically targeting large language model powered systems and interfaces, which demonstrate unprecedented capabilities including reasoning ~\cite{wei2022chain} and human-like interactions ~\cite{xi2023rise}, remains scarce. This is the first study we are aware of which looks at the impact of variation over multiple responses as a means of potentially reducing anthropomorphization and calibrating trust of a large language model powered systems and interfaces.

\subsection{Trust}
Previous research has demonstrated that when users anthropomorphize an AI system, it can significantly influence the level of trust they place in the system~\cite{hoff2015trust, kulms2019more}. For example, \citet{de2016almost} discovered that when users do not anthropomorphize AI agents, their trust in them decreases. 

The consequences of placing undue trust in AI systems can be significant, as evidenced by law enforcement disproportionately targeting marginalized and under-served communities due to embedded biases and ``bias-tainted data”~\cite{criminal-justice-ai, eubanks2018automating}, and the consequences of drivers overestimating the capabilities of self-driving features~\cite{autopilot-crashes}. To address this issue, researchers have advocated for the design of interfaces that assist users in calibrating their trust in the underlying system~\cite{lee2004trust, amershi2019guidelines}, thereby preventing them from over relying on it. According to \citet{lee2004trust}, how much a user trusts a system is inextricably linked to how much they rely on a software system, particularly when the inner workings of the system are unknown or unclear. 

Multiple interventions have been proposed to help users calibrate their trust in AI systems, including explanations~\cite{naiseh2021explainable} and confidence scores~\cite{zhang2020effect}. Our work extends this research by evaluating an intervention aimed at reducing anthropomorphization and increasing appropriate trust in the large language model they are interacting with. By presenting multiple responses generated by a single LLM in response to a single prompt, we aim to highlight that the model's behavior is a reflection of its sampling from a learned probability distribution, and each response represents a different sequence of words that the model considers plausible given the input prompt, thereby uncovering its probabilistic nature and what it does and does not consistently assert. We hypothesize that this becomes increasingly clear when the interface reveals the latent variation in the model's answers to the same prompt. 

\subsection{Measuring anthropomorphism and trust}
\subsubsection{Anthropomorphism measures}Waytz et al. proposes Individual Differences in Anthropomorphism Questionnaire (IDAQ), which captures individual differences in the potentially stable tendency to anthropomorphize~\cite{waytz2010sees}. This survey instrument has been widely adopted, and  contains questions about the consciousness, free will, mind, emotions, and intentions of various non-human entities. We use this questionnaire to measure our participants' general tendency to anthropomorphize.
To assess the extent to which users anthropomorphize the LLM-powered system used in this study, we utilize the anthropomorphization subsection of the Godspeed questionnaire~\cite{bartneck2009measurement}. This questionnaire, which is often used in human robot and agent interaction, includes criteria such as whether a response (in our case) is natural, machine-like, conscious, and/or lifelike. 

\subsubsection{Trust measures}  Several measures have been proposed to evaluate user trust in software systems, such as the Trust in Automation Scale~\cite{lee1992trust}, Human-Robot Interaction Trust Scale~\cite{yagoda2012you}, and Human-Computer Trust (HCT) scale developed by Madsen and Gregor~\cite{Madsen2000MeasuringHT}. In this paper, we employ the HCT scale due to its comprehensiveness and relevance to our research objectives. The HCT scale consists of five main constructs that capture the various facets of trust: perceived understandability, perceived technical competence, perceived reliability, personal attachment, and faith~\cite{Madsen2000MeasuringHT}. 
   


    

\section{User Study}
We conducted a within-subjects experiment with three conditions, each reflecting a different way of consuming generative AI answers to a given question. 

\subsection{Recruitment and Participants}
Nineteen participants were recruited at our university. Eight participants were between the ages of 18-24 and eleven were between the ages of 25-34. The participants were roughly gender balanced, with 11 women and 8 men. Eleven participants were PhD students, six were Masters students, one was an MD student, and one was a former Master's student who is still serving as a teaching assistant for a university course. 

\subsection{Protocol}
Each participant's session took approximately 50-60 minutes to complete.
Participants, after giving informed consent, completed a brief survey about their demographics and prior experiences with AI. This survey included the IDAQ~\cite{de2016almost} (see Appendix~\ref{app:idaq}) and NCS-6~\cite{cacioppo1982need} (see Appendix~\ref{app:ncs-6}) surveys, to capture their tendency to anthropomorphize and preference for engaging in cognitively demanding tasks, respectively. Participants were then assigned one of four possible questions; this question would be the same question they would contemplate answers to for the rest of the session. They were first asked to come up with their own answer to the question, then they answered the same question again in each of three conditions. Within each condition, participants were invited to talk aloud about their thought process. 

The four questions, which were uniformly assigned across participants, were chosen as examples of the various types of inquiries one might ask an LLM:
\begin{enumerate}
    \item \textit{Who invented the lightbulb?} [Factual, but with an answer that can be nuanced]
    \item \textit{``Keira, Clarissa, and Olive are not relatives. Keira and Clarissa have brothers, and Olive has a sister. Clarissa and Olive are the youngest in their families, and Keira is the oldest in hers. Who of them has an elder brother?'' \cite{logiclike2025riddles}} [Factual, logic-type]
    \item \textit{Is Jupiter more dense than Saturn? Why?} [Factual, with explanation requested]
    \item \textit{I am a senior engineer at a technology company. What is the one most effective way to maintain a healthy work-life balance? Be concise.} [Opinion, life advice]
\end{enumerate}

\begin{figure}[t]
    \centering
    \includegraphics[width=\linewidth]{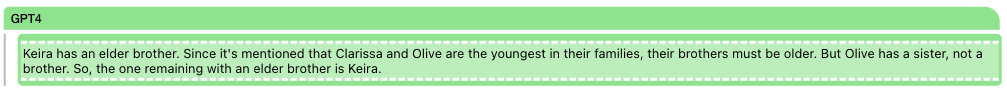}
    \caption{Single response}
    \label{fig:single}
\end{figure}
\begin{figure}[t]
    \centering
    \includegraphics[width=150mm, scale=5]{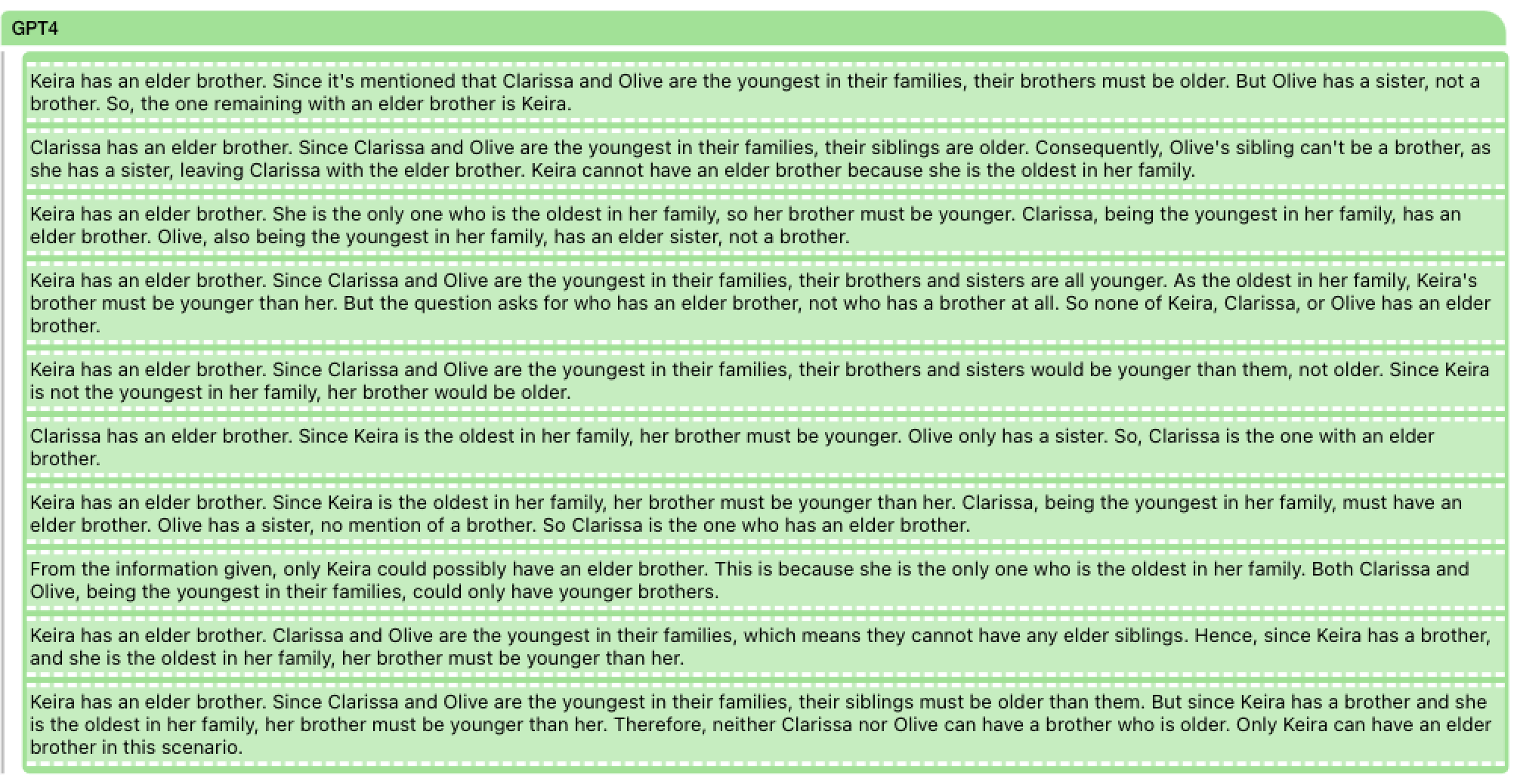}
    \caption{Ten responses without cognitive support}
    \label{fig:ten-responses-N}
\end{figure}
\begin{figure}[t]
    \centering
    \includegraphics[width=150mm, scale=5]{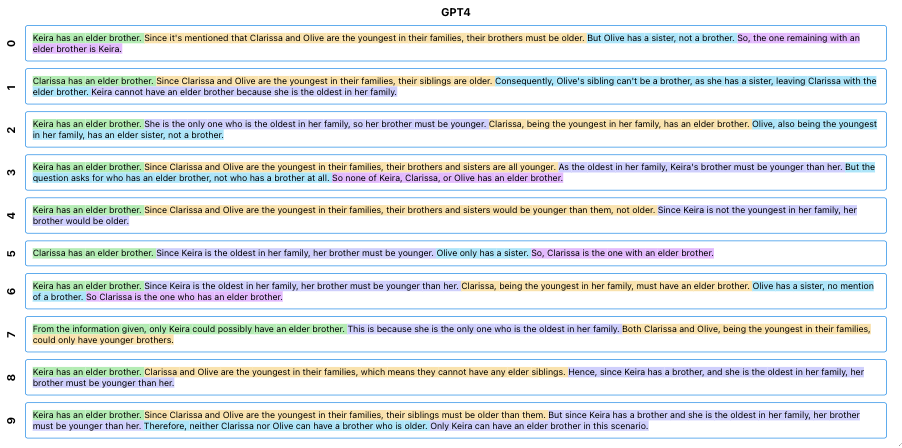}
    \caption{Ten responses with cognitive support}
    \label{fig:ten-responses-M}
\end{figure}

Using GPT-4 with a temperature setting of 1\footnote{The temperature setting was chosen to produce noticeable variation within just 10 LLM-generated responses, to cut down on tiresome repetition and expose inherent stochasticity. The default temperature setting of ChatGPT is unknown to us at this time, and the range of 0.8 -- 1.0 has been suggested by a popular LinkedIn blog for marketers~\cite{berkowitz_article} to be preferred ``when brainstorming innovative campaign ideas, crafting engaging social media content, or seeking fresh perspectives on a topic''.}, the three conditions were: (1) seeing a single LLM-generated answer to the question (\textit{Condition 1}) as shown in Figure.~\ref{fig:single}, (2) seeing ten LLM-generated answers to the question (with no cognitive support) (\textit{Condition N}) as shown in Figure.~\ref{fig:ten-responses-N}, and (3) seeing another set of ten LLM-generated answers with an interface feature that supports cognitively engaging with the answers (\textit{Condition M}) as shown in Figure.~\ref{fig:ten-responses-M}. The cognitive support mechanism was introduced to help users manage the cognitive load of seeing multiple responses. It uses Positional Diction Clustering (PDC), an algorithm that highlights structural and semantic similarities across multiple model responses in the same color~\cite{gero2024supporting}. When in a condition with multiple LLM responses shown at once, participants were told to ``imagine that you have asked the language model the same question on 10 separate occasions.'' This phrasing isolated the impact of seeing multiple responses on their perception of and reasoning about the LLM from the impact of the interface design itself. Participants were given up to two minutes to inspect a single generated response, up to four minutes to inspect ten responses with cognitive support, and up to four minutes to inspect ten responses without cognitive support.  

After each condition, the participant answered the NASA-TLX~\cite{hart1988development} questions about the process of answering the question in that condition, the HCT~\cite{madsen2000measuring} questions (see Appendix~\ref{app:hct}) about trust in the AI, and the Godspeed~\cite{bartneck2009measurement} questions (see Appendix~\ref{app:godspeed}) about anthropomorphization of the AI.

The counterbalancing of conditions had two levels. At the top level, half the participants experienced the 1-answer condition first and then both conditions with 10 LLM-generated answers, sequentially, after. The other half of participants experienced both conditions with 10 LLM-generated answers, sequentially, first, and then experienced the 1-answer condition. At the second level, half the participants experienced the condition with 10 LLM-generated answers without cognitive support first and half the participants experienced the condition with 10 LLM-generated answers with cognitive support first. We designed the experiment this way to collect as much data as possible on the impact of 10 vs 1 answer, with additional detail within that experimental design for the impact of cognitive support for engaging with those 10 answers.
After completing all three conditions, participants were invited to participate in a semi-structured interview about their experiences in each condition. 


\subsection{Qualitative Analysis}
Qualitative data for this study consists of notes collected as participants thought aloud while completing their tasks, post-condition surveys, and participants' answers in a post study semi-structured interview. To conduct the qualitative analysis, two researchers followed the general inductive approach defined by Thomas ~\cite{thomas2006general}. The 19 research studies were divided between two researchers. To ensure that both researchers had context of all studies, both researchers read the transcripts for all studies. The authors pulled quotes from all transcripts and then independently coded >50\% of the pulled quotes. They then created an codebook based on the shared codes and themes that were identified. Researchers used this initial codebook to code the quotes of 10 participants. They discussed disagreements, established inter-rater reliability (IRR), and revised the codebook. They used the revised codebook to recode the quotes of 10 participants and repeat the thematic analysis process. Finally, both participants coded quotes from the remaining 9 participants and reached IRR.

\subsection{Quantitative Analysis}

We continued our analysis by exploring whether question type impacted participant's experience with the model responses. For each question type (within each condition), we explored whether there were differences in how participants responded to the post task survey questions (perceived workload, human-computer trust, and anthropomorphism). For each question type and scale, we conducted a series of one-way ANOVA omnibus tests, followed by Mann-Whitney U tests.


\section{Results}
\subsection{Factual Question}
Five participants (P1, P3, P4, P5, and P17) answered the factual question, ``Who invented the lightbulb?" with the assistance of three language-model-powered interfaces. The study found that showing the users multiple responses allowed them to perform cross-checks across responses. The consistency between the generated responses served, for some participants, as an indicator of the accuracy of the language model---or a proxy of its ability to accurately answer the question. For example, P5 stated, "It’s useful to have 10 different outputs for the same prompt so that I can have a little bit more confirmation." According to P17, the ability to perform cross-checks could be particularly useful when the language model is prompted to answer a less common question: "If I'm looking for something that's pretty rare, I would probably want to verify the output, or if the 10 generations in the second condition [where multiple responses are shown] do not agree with each other very often, that would decrease my trust in the system output." Similarly, consistency between the generated responses suggested that the language model was reliable. As P4 stated, ``The ten responses I saw are all pretty similar, with a lot of overlap, which makes me feel like it's pretty reliable." 
Interestingly, P17 argued that multiple differing responses gave them a sense that the model was knowledgeable, noting ``For the second condition [\textit{Condition N}], I think it's providing more facts, whether some of the facts are true or not, but it definitely shows it's very capable of giving out good information. And for the third one [\textit{Condition M}], the study shows the system has the ability to rephrase a lot of things and also dig out less-known facts. So, I think generally, it's really knowledgeable."

\subsection{Logic Question}
Five participants (P2, P6, P7, P8, and P18) answered the logical question about Keira, Clarissa, and Olive with the assistance of three different language-model-powered interfaces. Multiple participants (P6, P7, P8, and P17) pointed out differences between the generated responses, including the final answer and intermediate reasoning steps. For example, P8 remarked, “I was just surprised that 70\% of the time, it was not able to have the same answer.” This had its pros. For example, P6 explained that it exposed information about the underlying system: “I found it interesting to see the diverse responses. Some were just plainly incorrect, so it was intriguing to see how it processes logic, which was something I was looking for. It can connect logical pieces to conclude, but it doesn't do it consistently.” Additionally, P7 mentioned that it helped them calibrate their trust in the model, stating, “If it's a single one, I'm more inclined to trust it because I don't know any better. I would have to think hard, versus if I see a bunch of different answers, I think, okay, it has no idea what it's talking about.” 

However, there were also cons. P8 explained that differing responses prevented them from predicting the model's behavior, stating, ``Do I know what will happen the next time I use the system? Absolutely not." Moreover, the inconsistency between the generated responses caused participants (P2, P8, and P18) to question the model's ability to answer questions requiring logical reasoning. For example, P2 stated, “The times I've used it... it's good to provide you with information. But... it's not that good in giving you sort of answers on analytical tasks.” P2 further elaborated on the model’s limitations for logical tasks by stating, “Its logic power is poor, but it's good at giving you information, summarizing information.” P18 offered a possible explanation for this: limitations in the input data, the base model's data capacity, and the configuration of its parameters.

Furthermore, P7 and P18 expressed contrasting views on the human-likeness of the language model's responses. P7 criticized the model's output as lacking genuine thought, describing it as ``word salad." Conversely, P18 stated: “How would you rate responses from artificial to lifelike? Very lifelike, so a five, with all these different responses.”

\subsection{Factual Question with Justification}

Five participants (P9, P10, P11, P12, and P19) answered the factual question with justification, "Is Jupiter more dense than Saturn? Why?" with the assistance of three different language-model-powered interfaces. The participants reported differences between the generated responses. In particular, P9 and P19 mentioned that the “structure” of the overall response and “focus” of the justification part of responses differed. As P9 stated: "Even though the answers are correct each time, the structure and focus differ." However, since the “core” or “main” part of the answer was consistent some of the participants (P12, P19) expressed that their trust in the system did not decrease. For example, P19 went so far as to say that: “If I'm asking the same question and I'm always getting the same main answer, I wouldn't even check Google. I would assume that it's correct.” Similarly, P12 stated: “My perception changed after receiving the same response ten times. This consistency made me feel that the model is very confident in its answers, suggesting it might be drawing from a reliable source or database. Therefore, I trust it.”

\subsection{Advice}
Four participants (P13, P14, P15, P16) answered the advise-based question about  work-life balance with the assistance of three different language-model-powered interfaces. Participants noted that while the language model’s responses were consistent in their main thesis---particularly around the theme of setting boundaries---there was a lack of specificity in the advice as well as variation in the answers. For example, P13 stated: "I feel like the crux of this, the way that the LLM is responding or suggesting I think that's good. But I think that it's also a bit vague as well... it feels like a shorter answer than I would like." P13 added, however, that it does cover all the “core points” and “elements” for someone to think about it. Similarly, P15 noted that “But every like first sentence that it would produce, it was so surprising how it was just very, very general almost said the same exact thing.” Additionally, P14 observed that the model "kind of missed different directions" such as delegating tasks. 
The level of detail in the responses was perceived as a double-edged sword. On one hand, the amount of detail made the responses seem less human-like, as if there was a machine behind them. As P14 stated, “the amount of detail made it less human-like, I think... But the answers themselves, I think, were quite human-like". On the other hand, P14 felt that the responses that provided more specific suggestions like digital detox and physical activities were more human-like and resembled the advice one might get from a friend or advisor. They expressed, “I think the fact that the last one gave sort of more detailed responses made it more human-like, I think because you know, I guess normally you would maybe ask that like to a friend or maybe some sort of advisor or at least any would want a more specific response”.

Participants expressed that variation in response could make the system seem more human-like, as it reflects the natural variation in human answers. For example, P16 stated, "If they provide more responses, that could be more like a human… if you ask a person 10 times, the main thing might be the same, but the human will respond with different answers." Moreover, P16 added that the ability to view multiple responses increases their trust, as it allows them to find information that they personally approve of, stating, “I tend to trust the more things. I can always find some information that I really approve of. But if we only see one piece of information [I might just] disapprove [of] it.” However, having access to multiple responses revealed conflicting advice regarding work-life balance strategies. As P15 stated, “There seems to be some conflicting advice with how the person should go about their work-life balance? Like do you strictly adhere to setting work and personal hours? Do you increase flexibility? Do you focus more on physical exercise?...Or do you focus more on personal interests, and mental well-being? And I mean, I know those aren't mutually exclusive, but it just seemed like it wasn't giving extremely consistent advice.”

\subsection{Quantitative Results}
To evaluate the impact of our conditions on participants answers to the NASA-TLX, HCT Scale, and Godspeed questionnaire across the three conditions we performed an omnibus statistical test, i.e., an omnibus ANOVA analysis across all data, i.e., all three conditions (engaging with one response, engaging with 10 responses with cognitive support, and engaging with 10 responses without cognitive support). This omnibus test did not reveal a significant effect of condition on participants answers to any of these survey instruments.

Despite this, to gather (possibly faulty) information scent to shape future larger studies, we ran further statistical tests on different subparts of our study results. With that caveat, we report p-values of those follow-up tests.

We evaluated whether there were statistically significant differences in how participants responded to the survey instruments, across the three conditions, for each question type separately and some distinct subcomponents of questionnaires separately. Across the three conditions, there was no significant difference detected for the perceived workload across the question types. 

For \textit{Condition N}, ANOVA revealed that the variance in \textit{perceived reliability} of the model across different question types had some lower p-values. For example, participants who answered the factual question (with justification) and the advice question differed on the system always providing information needed to make a decision (p=0.03). Participants who answered the factual question and factual question with justification had different perceptions of the system's ability to perform reliably. As seen in the qualitative data, this was likely a result of the justification varying across responses. 

We ran Mann-Whitney U tests on differences in each of the subsections of the HCT (e.g., perceived technical competence, reliability, understandability, etc.) across question type. For example, participants who engaged with responses to the logic question and the advice question differed in their perception of the system being able to accurately use information (p=0.02); participants who reviewed advice responses noted that the model did not incorporate details given to it. The p-value for the difference in participants' perceived understandability of the model across the question types was also relatively low. The responses given for the logic question and the factual question with justification resulted in differences in understanding how to manage the system to retrieve desired information (p=0.02). Participants who answered the factual question and logic question had conflicting levels of attachment to using the system (p=0.04). The qualitative data showed that participants noticed that the model was consistent in its answers to the factual question, but there were different, sometimes conflicting, responses to the logical based question.
For additional details on differences calculated across question type, see Appendix~\ref{app:quantitiave-results}.


\section{Discussion and Future Work}

Our study sought to investigate the impact of two design interventions: (1) the impact of displaying multiple LLM responses on user perceptions of the system's trustworthiness, reliability, and anthropomorphic qualities and (2) the impact of cognitive support 
for engaging with multiple responses. 


\subsection{Impact on User Perceptions}

We found no statistically significant effect regarding the impact of displaying multiple LLM responses on user perceptions of the system. However, interviews conducted after participants responded to questions across all three conditions revealed challenges in assessing the model's reliability and calibrating their trust, particularly when they were only shown a single response (see Section~\ref{sec:unabletojudge}).

Our qualitative analysis found that showing multiple responses simultaneously allowed users to perform cross-"verification" by assessing the consistency among multiple responses. This form of internal "validation" can be particularly useful when users provide rare or complex queries. The results suggest that this additional information helped users feel like they could make a more informed guess about the system’s capabilities, which they felt they could not do when they were only presented with one response. However, we argue that future work should consider the potential limitations of using consistency as a sole proxy for accuracy or reliability. For example, what if the wrong answer is generated more than the right answer? 

Interestingly, showing multiple responses simultaneously that disagreed did not always serve as a bad sign. Instead, one participant saw the system as more knowledgeable. Moreover, another described the variation as more human-like. This suggests that variability in responses, rather than diminishing trust, can enhance the perceived depth and breadth of the LLM’s knowledge. This aspect could be further explored to understand how variability influences user perceptions of LLMs' capabilities and their anthropomorphization of the LLM. Furthermore, this disagreement was more welcomed in the advice-based question group by the participants. This suggests that users may value diverse perspectives when seeking guidance on subjective matters. This could inform the design of LLM interfaces for advice-giving, where presenting a range of suggestions could be more beneficial than offering a single, definitive answer.


\subsection{Cognitive Load and Interface Design}

Our findings indicate that cognitive support can assist users in more efficiently parsing through multiple outputs, but this was not ubiquitous. 
If someone truly engages with multiple LLM responses instead of one, it is impossible for the cognitive load to be the same. In this study, however, any increase in the cognitive load between the one and many LLM response conditions was not statistically significant; participants may not have been thoroughly engaging enough with the multiple responses to increase in their NASA-TLX scores. Or the magnitude of the difference in cognitive load of fully engaging with multiple LLM responses may not have been sufficient to register as significant. 

Cognitive support, if well designed, may be able to ameliorate any increase in cognitive load through pre-computation of similarities and differences and interface design that makes the meaning of these pre-computed qualities clear rather than confusing (and therefore generating their own increase in cognitive load). We chose PDC-based similar sentence highlighting from~\cite{gero2024supporting} due to its intended alignment with these design goals. 
Regardless, the cognitive support deployed in this experiment, which is one of many possible such interventions, may have, on average, added as much cognitive load as it alleviated, or it may have had a positive impact that was not measurable for the same reason that the difference between the cognitive load of receiving one and many LLM responses was not measurable. We believe the design space of cognitive supports for consuming multiple LLM responses should be explored further, given the potential benefits to LLM users of consuming multiple responses to the same query, e.g., ability to judge reliability.

\subsection{Future work}

\subsubsection{Disaggregating Data based on Psychological Traits}

Our study did not disaggregate data based on psychological traits, i.e., Need for Cognition (NFC) or Individual Differences in Anthropomorphism Questionnaire (IDAQ) scores. In particular, we found that IDAQ scores were low on average (see Figure.~\ref{fig:idaq}), indicating lower tendencies to attribute human-like qualities to non-human entities. This is relevant because, if participants start out with little anthropomorphization of the system, any intervention designed to further lower it may be more difficult to detect. Future studies with larger sample sizes could collect more participants with higher IDAQ scores and perform such disaggregation to examine how these traits might influence system anthropomophization and perceptions of its trustworthiness and reliability. Lastly, our participant pool was heavily skewed towards individuals within computer science degree programs. 
Recruiting more diverse participants would help disambiguate whether (1) tendency to anthropomorphize and (2) familiarity with the underlying technology mediates the impacts of the interventions we investigated.

\subsubsection{Investigating Cognitive Load Metrics Further}

The absence of statistical significance related to cognitive load metrics in our study raises important questions and opens avenues for further research. The observed lack of difference in cognitive load when users were presented with multiple responses versus one could be attributed to several factors, including insufficient statistical power. This suggests that the study may not have had enough participants or was not designed sensitively enough to detect a true effect. Additionally, the inherent similarity of the generated responses might not have sufficiently challenged participants' cognitive processing, leading to the observed lack of significant findings. 

In addition to considering individuals' subjective preferences and answers to the survey instruments, our next study will utilize eye tracking technology to analyze participants' reading behavior and identify how, or if, their gaze transitions from analogous segments in one response to those in another. Detecting instances where participants perceive discrepancies between answers, and whether the interface intervention for cognitive support assists with that, could offer valuable insights into subtle interface modifications that allow people to take in redundant but possible inconsistent information. 


\subsubsection{Exploring Question Types and Response Consistency}



Future work could explore how the question type —whether it is factual or opinion based— can influence the consistency of the LLM’s responses, and thus how users perceive and trust these responses. For instance, LLMs might show high consistency in answering factual questions due to their reliance on retrievable data (e.g., Retrieval-Augmented Generation), but might vary more in responses to opinion-based questions, which are subjective and can be influenced by the training data's diversity. This variability could affect users' trust differently, as they might not expect—or desire—the same level of consistency in answers to these more subjective queries.

\section*{Acknowledgments}
This material is based upon work supported by the National Science Foundation under Grants No. IIS-2107391 and IIS-2040880. This paper was accepted to the Trust and Reliance in Evolving Human-AI Workflows (TREW) Workshop at CHI 2024. We thank the paper reviewers and workshop attendees for their valuable and insightful feedback.

\bibliographystyle{ACM-Reference-Format}
\bibliography{main}


\begin{thebibliography}{41}


\ifx \showCODEN    \undefined \def \showCODEN     #1{\unskip}     \fi
\ifx \showDOI      \undefined \def \showDOI       #1{#1}\fi
\ifx \showISBNx    \undefined \def \showISBNx     #1{\unskip}     \fi
\ifx \showISBNxiii \undefined \def \showISBNxiii  #1{\unskip}     \fi
\ifx \showISSN     \undefined \def \showISSN      #1{\unskip}     \fi
\ifx \showLCCN     \undefined \def \showLCCN      #1{\unskip}     \fi
\ifx \shownote     \undefined \def \shownote      #1{#1}          \fi
\ifx \showarticletitle \undefined \def \showarticletitle #1{#1}   \fi
\ifx \showURL      \undefined \def \showURL       {\relax}        \fi
\providecommand\bibfield[2]{#2}
\providecommand\bibinfo[2]{#2}
\providecommand\natexlab[1]{#1}
\providecommand\showeprint[2][]{arXiv:#2}

\bibitem[Amershi et~al\mbox{.}(2019)]%
        {amershi2019guidelines}
\bibfield{author}{\bibinfo{person}{Saleema Amershi}, \bibinfo{person}{Dan Weld}, \bibinfo{person}{Mihaela Vorvoreanu}, \bibinfo{person}{Adam Fourney}, \bibinfo{person}{Besmira Nushi}, \bibinfo{person}{Penny Collisson}, \bibinfo{person}{Jina Suh}, \bibinfo{person}{Shamsi Iqbal}, \bibinfo{person}{Paul~N Bennett}, \bibinfo{person}{Kori Inkpen}, {et~al\mbox{.}}} \bibinfo{year}{2019}\natexlab{}.
\newblock \showarticletitle{Guidelines for human-AI interaction}. In \bibinfo{booktitle}{\emph{Proceedings of the 2019 chi conference on human factors in computing systems}}. \bibinfo{pages}{1--13}.
\newblock


\bibitem[Bartneck et~al\mbox{.}(2009)]%
        {bartneck2009measurement}
\bibfield{author}{\bibinfo{person}{Christoph Bartneck}, \bibinfo{person}{Dana Kuli{\'c}}, \bibinfo{person}{Elizabeth Croft}, {and} \bibinfo{person}{Susana Zoghbi}.} \bibinfo{year}{2009}\natexlab{}.
\newblock \showarticletitle{Measurement instruments for the anthropomorphism, animacy, likeability, perceived intelligence, and perceived safety of robots}.
\newblock \bibinfo{journal}{\emph{International journal of social robotics}}  \bibinfo{volume}{1} (\bibinfo{year}{2009}), \bibinfo{pages}{71--81}.
\newblock


\bibitem[Berkowitz(2023)]%
        {berkowitz_article}
\bibfield{author}{\bibinfo{person}{David Berkowitz}.} \bibinfo{year}{2023}\natexlab{}.
\newblock \bibinfo{title}{Temperature Check: A Guide to the Best ChatGPT Feature You're (Probably) Not Using}.
\newblock
\newblock
\urldef\tempurl%
\url{https://www.linkedin.com/pulse/temperature-check-guide-best-chatgpt-feature-youre-using-berkowitz/}
\showURL{%
\tempurl}
\newblock
\shownote{Accessed on: 9 October 2023}.


\bibitem[Bi and Huang(2023)]%
        {bi2023create}
\bibfield{author}{\bibinfo{person}{Nanyi Bi} {and} \bibinfo{person}{Janet Yi-Ching Huang}.} \bibinfo{year}{2023}\natexlab{}.
\newblock \showarticletitle{I create, therefore I agree: Exploring the effect of AI anthropomorphism on human decision-making}. In \bibinfo{booktitle}{\emph{Companion Publication of the 2023 Conference on Computer Supported Cooperative Work and Social Computing}}. \bibinfo{pages}{241--244}.
\newblock


\bibitem[Bu{\c{c}}inca et~al\mbox{.}(2021)]%
        {tothinkortrust}
\bibfield{author}{\bibinfo{person}{Zana Bu{\c{c}}inca}, \bibinfo{person}{Maja~Barbara Malaya}, {and} \bibinfo{person}{Krzysztof~Z Gajos}.} \bibinfo{year}{2021}\natexlab{}.
\newblock \showarticletitle{To trust or to think: cognitive forcing functions can reduce overreliance on AI in AI-assisted decision-making}.
\newblock \bibinfo{journal}{\emph{Proceedings of the ACM on Human-Computer Interaction}} \bibinfo{volume}{5}, \bibinfo{number}{CSCW1} (\bibinfo{year}{2021}), \bibinfo{pages}{1--21}.
\newblock


\bibitem[Cacioppo and Petty(1982)]%
        {cacioppo1982need}
\bibfield{author}{\bibinfo{person}{John~T Cacioppo} {and} \bibinfo{person}{Richard~E Petty}.} \bibinfo{year}{1982}\natexlab{}.
\newblock \showarticletitle{The need for cognition.}
\newblock \bibinfo{journal}{\emph{Journal of personality and social psychology}} \bibinfo{volume}{42}, \bibinfo{number}{1} (\bibinfo{year}{1982}), \bibinfo{pages}{116}.
\newblock


\bibitem[De~Visser et~al\mbox{.}(2016)]%
        {de2016almost}
\bibfield{author}{\bibinfo{person}{Ewart~J De~Visser}, \bibinfo{person}{Samuel~S Monfort}, \bibinfo{person}{Ryan McKendrick}, \bibinfo{person}{Melissa~AB Smith}, \bibinfo{person}{Patrick~E McKnight}, \bibinfo{person}{Frank Krueger}, {and} \bibinfo{person}{Raja Parasuraman}.} \bibinfo{year}{2016}\natexlab{}.
\newblock \showarticletitle{Almost human: Anthropomorphism increases trust resilience in cognitive agents.}
\newblock \bibinfo{journal}{\emph{Journal of Experimental Psychology: Applied}} \bibinfo{volume}{22}, \bibinfo{number}{3} (\bibinfo{year}{2016}), \bibinfo{pages}{331}.
\newblock


\bibitem[Eubanks(2018)]%
        {eubanks2018automating}
\bibfield{author}{\bibinfo{person}{Virginia Eubanks}.} \bibinfo{year}{2018}\natexlab{}.
\newblock \bibinfo{booktitle}{\emph{Automating inequality: How high-tech tools profile, police, and punish the poor}}.
\newblock \bibinfo{publisher}{St. Martin's Press}.
\newblock


\bibitem[Eyssel et~al\mbox{.}(2012)]%
        {eyssel2012if}
\bibfield{author}{\bibinfo{person}{Friederike Eyssel}, \bibinfo{person}{Dieta Kuchenbrandt}, \bibinfo{person}{Simon Bobinger}, \bibinfo{person}{Laura De~Ruiter}, {and} \bibinfo{person}{Frank Hegel}.} \bibinfo{year}{2012}\natexlab{}.
\newblock \showarticletitle{'If you sound like me, you must be more human' on the interplay of robot and user features on human-robot acceptance and anthropomorphism}. In \bibinfo{booktitle}{\emph{Proceedings of the seventh annual ACM/IEEE international conference on Human-Robot Interaction}}. \bibinfo{pages}{125--126}.
\newblock


\bibitem[Feine et~al\mbox{.}(2019)]%
        {feine2019taxonomy}
\bibfield{author}{\bibinfo{person}{Jasper Feine}, \bibinfo{person}{Ulrich Gnewuch}, \bibinfo{person}{Stefan Morana}, {and} \bibinfo{person}{Alexander Maedche}.} \bibinfo{year}{2019}\natexlab{}.
\newblock \showarticletitle{A taxonomy of social cues for conversational agents}.
\newblock \bibinfo{journal}{\emph{International Journal of Human-Computer Studies}}  \bibinfo{volume}{132} (\bibinfo{year}{2019}), \bibinfo{pages}{138--161}.
\newblock


\bibitem[Gero et~al\mbox{.}(2024)]%
        {gero2024supporting}
\bibfield{author}{\bibinfo{person}{Katy~Ilonka Gero}, \bibinfo{person}{Chelse Swoopes}, \bibinfo{person}{Ziwei Gu}, \bibinfo{person}{Jonathan~K. Kummerfeld}, {and} \bibinfo{person}{Elena~L. Glassman}.} \bibinfo{year}{2024}\natexlab{}.
\newblock \bibinfo{title}{Supporting Sensemaking of Large Language Model Outputs at Scale}.
\newblock
\newblock
\showeprint[arxiv]{2401.13726}~[cs.HC]


\bibitem[Hao(2019)]%
        {criminal-justice-ai}
\bibfield{author}{\bibinfo{person}{Karen Hao}.} \bibinfo{year}{2019}\natexlab{}.
\newblock \showarticletitle{AI is sending people to jail—and getting it wrong}.
\newblock \bibinfo{howpublished}{\url{https://www.technologyreview.com/2019/01/21/137783/algorithms-criminal-justice-ai/}}.
\newblock  (\bibinfo{year}{2019}).
\newblock


\bibitem[Hart and Staveland(1988)]%
        {hart1988development}
\bibfield{author}{\bibinfo{person}{Sandra~G Hart} {and} \bibinfo{person}{Lowell~E Staveland}.} \bibinfo{year}{1988}\natexlab{}.
\newblock \showarticletitle{Development of NASA-TLX (Task Load Index): Results of empirical and theoretical research}.
\newblock In \bibinfo{booktitle}{\emph{Advances in psychology}}. Vol.~\bibinfo{volume}{52}. \bibinfo{publisher}{Elsevier}, \bibinfo{pages}{139--183}.
\newblock


\bibitem[Hoff and Bashir(2015)]%
        {hoff2015trust}
\bibfield{author}{\bibinfo{person}{Kevin~Anthony Hoff} {and} \bibinfo{person}{Masooda Bashir}.} \bibinfo{year}{2015}\natexlab{}.
\newblock \showarticletitle{Trust in automation: Integrating empirical evidence on factors that influence trust}.
\newblock \bibinfo{journal}{\emph{Human factors}} \bibinfo{volume}{57}, \bibinfo{number}{3} (\bibinfo{year}{2015}), \bibinfo{pages}{407--434}.
\newblock


\bibitem[Jensen et~al\mbox{.}(2021)]%
        {jensen2021trust}
\bibfield{author}{\bibinfo{person}{Theodore Jensen}, \bibinfo{person}{Mohammad Maifi~Hasan Khan}, \bibinfo{person}{Md~Abdullah~Al Fahim}, {and} \bibinfo{person}{Yusuf Albayram}.} \bibinfo{year}{2021}\natexlab{}.
\newblock \showarticletitle{Trust and anthropomorphism in tandem: the interrelated nature of automated agent appearance and reliability in trustworthiness perceptions}. In \bibinfo{booktitle}{\emph{Designing interactive systems conference 2021}}. \bibinfo{pages}{1470--1480}.
\newblock


\bibitem[Kimp({[n.\,d.]})]%
        {bard}
\bibfield{author}{\bibinfo{person}{Kimp}.} \bibinfo{year}{[n.\,d.]}\natexlab{}.
\newblock \bibinfo{title}{Google Bard: Everything You Need To Know About The ChatGPT Alternative}.
\newblock
\newblock
\urldef\tempurl%
\url{https://www.kimp.io/google-bard/}
\showURL{%
\tempurl}
\newblock
\shownote{Accessed on: 9 October 2023}.


\bibitem[Kulms and Kopp(2019)]%
        {kulms2019more}
\bibfield{author}{\bibinfo{person}{Philipp Kulms} {and} \bibinfo{person}{Stefan Kopp}.} \bibinfo{year}{2019}\natexlab{}.
\newblock \showarticletitle{More human-likeness, more trust? The effect of anthropomorphism on self-reported and behavioral trust in continued and interdependent human-agent cooperation}.
\newblock In \bibinfo{booktitle}{\emph{Proceedings of mensch und computer 2019}}. \bibinfo{pages}{31--42}.
\newblock


\bibitem[Laban(2021)]%
        {laban2021perceptions}
\bibfield{author}{\bibinfo{person}{Guy Laban}.} \bibinfo{year}{2021}\natexlab{}.
\newblock \showarticletitle{Perceptions of anthropomorphism in a chatbot dialogue: the role of animacy and intelligence}. In \bibinfo{booktitle}{\emph{Proceedings of the 9th International Conference on Human-Agent Interaction}}. \bibinfo{pages}{305--310}.
\newblock


\bibitem[Lee and Moray(1992)]%
        {lee1992trust}
\bibfield{author}{\bibinfo{person}{John Lee} {and} \bibinfo{person}{Neville Moray}.} \bibinfo{year}{1992}\natexlab{}.
\newblock \showarticletitle{Trust, control strategies and allocation of function in human-machine systems}.
\newblock \bibinfo{journal}{\emph{Ergonomics}} \bibinfo{volume}{35}, \bibinfo{number}{10} (\bibinfo{year}{1992}), \bibinfo{pages}{1243--1270}.
\newblock


\bibitem[Lee and See(2004)]%
        {lee2004trust}
\bibfield{author}{\bibinfo{person}{John~D Lee} {and} \bibinfo{person}{Katrina~A See}.} \bibinfo{year}{2004}\natexlab{}.
\newblock \showarticletitle{Trust in automation: Designing for appropriate reliance}.
\newblock \bibinfo{journal}{\emph{Human factors}} \bibinfo{volume}{46}, \bibinfo{number}{1} (\bibinfo{year}{2004}), \bibinfo{pages}{50--80}.
\newblock


\bibitem[Lee et~al\mbox{.}(2020)]%
        {lee2020perceiving}
\bibfield{author}{\bibinfo{person}{Sangwon Lee}, \bibinfo{person}{Naeun Lee}, {and} \bibinfo{person}{Young~June Sah}.} \bibinfo{year}{2020}\natexlab{}.
\newblock \showarticletitle{Perceiving a mind in a chatbot: effect of mind perception and social cues on co-presence, closeness, and intention to use}.
\newblock \bibinfo{journal}{\emph{International Journal of Human--Computer Interaction}} \bibinfo{volume}{36}, \bibinfo{number}{10} (\bibinfo{year}{2020}), \bibinfo{pages}{930--940}.
\newblock


\bibitem[LogicLike(2025)]%
        {logiclike2025riddles}
\bibfield{author}{\bibinfo{person}{LogicLike}.} \bibinfo{year}{2025}\natexlab{}.
\newblock \bibinfo{title}{20+ Logic Riddles and Good Riddle Questions with Answers}.
\newblock
\newblock
\urldef\tempurl%
\url{https://logiclike.com/en/logic-riddles-questions}
\showURL{%
\tempurl}
\newblock
\shownote{Accessed: 2024-02}.


\bibitem[Madsen and Gregor(2000a)]%
        {Madsen2000MeasuringHT}
\bibfield{author}{\bibinfo{person}{Maria Madsen} {and} \bibinfo{person}{Shirley Gregor}.} \bibinfo{year}{2000}\natexlab{a}.
\newblock \showarticletitle{Measuring human-computer trust}. In \bibinfo{booktitle}{\emph{11th australasian conference on information systems}}, Vol.~\bibinfo{volume}{53}. Citeseer.
\newblock


\bibitem[Madsen and Gregor(2000b)]%
        {madsen2000measuring}
\bibfield{author}{\bibinfo{person}{Maria Madsen} {and} \bibinfo{person}{Shirley Gregor}.} \bibinfo{year}{2000}\natexlab{b}.
\newblock \showarticletitle{Measuring human-computer trust}. In \bibinfo{booktitle}{\emph{11th australasian conference on information systems}}, Vol.~\bibinfo{volume}{53}. Citeseer, \bibinfo{pages}{6--8}.
\newblock


\bibitem[Metz(2023)]%
        {nyt-ai-human-reasoning}
\bibfield{author}{\bibinfo{person}{Cade Metz}.} \bibinfo{year}{2023}\natexlab{}.
\newblock \showarticletitle{Microsoft Says New A.I. Shows Signs of Human Reasoning}.
\newblock \bibinfo{howpublished}{\url{https://www.nytimes.com/2023/05/16/technology/microsoft-ai-human-reasoning.html}}.
\newblock  (\bibinfo{year}{2023}).
\newblock


\bibitem[Naiseh et~al\mbox{.}(2021)]%
        {naiseh2021explainable}
\bibfield{author}{\bibinfo{person}{Mohammad Naiseh}, \bibinfo{person}{Dena Al-Thani}, \bibinfo{person}{Nan Jiang}, {and} \bibinfo{person}{Raian Ali}.} \bibinfo{year}{2021}\natexlab{}.
\newblock \showarticletitle{Explainable recommendation: when design meets trust calibration}.
\newblock \bibinfo{journal}{\emph{World Wide Web}} \bibinfo{volume}{24}, \bibinfo{number}{5} (\bibinfo{year}{2021}), \bibinfo{pages}{1857--1884}.
\newblock


\bibitem[Nass et~al\mbox{.}(1993)]%
        {nass1993anthropomorphism}
\bibfield{author}{\bibinfo{person}{Clifford Nass}, \bibinfo{person}{Jonathan Steuer}, \bibinfo{person}{Ellen Tauber}, {and} \bibinfo{person}{Heidi Reeder}.} \bibinfo{year}{1993}\natexlab{}.
\newblock \showarticletitle{Anthropomorphism, agency, and ethopoeia: computers as social actors}. In \bibinfo{booktitle}{\emph{INTERACT'93 and CHI'93 conference companion on Human factors in computing systems}}. \bibinfo{pages}{111--112}.
\newblock


\bibitem[OpenAI(2023)]%
        {chatgpt}
\bibfield{author}{\bibinfo{person}{OpenAI}.} \bibinfo{year}{Accessed 2023}\natexlab{}.
\newblock \bibinfo{title}{ChatGPT}.
\newblock \bibinfo{howpublished}{\url{https://chat.openai.com/}}.
\newblock


\bibitem[Perplexity(2023)]%
        {perplexity-ai}
\bibfield{author}{\bibinfo{person}{Perplexity}.} \bibinfo{year}{Accessed 2023}\natexlab{}.
\newblock \bibinfo{title}{perplexity}.
\newblock \bibinfo{howpublished}{\url{https://www.perplexity.ai/}}.
\newblock


\bibitem[Riek et~al\mbox{.}(2009)]%
        {riek2009anthropomorphism}
\bibfield{author}{\bibinfo{person}{Laurel~D Riek}, \bibinfo{person}{Tal-Chen Rabinowitch}, \bibinfo{person}{Bhismadev Chakrabarti}, {and} \bibinfo{person}{Peter Robinson}.} \bibinfo{year}{2009}\natexlab{}.
\newblock \showarticletitle{How anthropomorphism affects empathy toward robots}. In \bibinfo{booktitle}{\emph{Proceedings of the 4th ACM/IEEE international conference on Human robot interaction}}. \bibinfo{pages}{245--246}.
\newblock


\bibitem[Schanke et~al\mbox{.}(2021)]%
        {schanke2021estimating}
\bibfield{author}{\bibinfo{person}{Scott Schanke}, \bibinfo{person}{Gordon Burtch}, {and} \bibinfo{person}{Gautam Ray}.} \bibinfo{year}{2021}\natexlab{}.
\newblock \showarticletitle{Estimating the impact of “humanizing” customer service chatbots}.
\newblock \bibinfo{journal}{\emph{Information Systems Research}} \bibinfo{volume}{32}, \bibinfo{number}{3} (\bibinfo{year}{2021}), \bibinfo{pages}{736--751}.
\newblock


\bibitem[Sheehan et~al\mbox{.}(2020)]%
        {sheehan2020customer}
\bibfield{author}{\bibinfo{person}{Ben Sheehan}, \bibinfo{person}{Hyun~Seung Jin}, {and} \bibinfo{person}{Udo Gottlieb}.} \bibinfo{year}{2020}\natexlab{}.
\newblock \showarticletitle{Customer service chatbots: Anthropomorphism and adoption}.
\newblock \bibinfo{journal}{\emph{Journal of Business Research}}  \bibinfo{volume}{115} (\bibinfo{year}{2020}), \bibinfo{pages}{14--24}.
\newblock


\bibitem[Siddiqui and Merrill(2023)]%
        {autopilot-crashes}
\bibfield{author}{\bibinfo{person}{Faiz Siddiqui} {and} \bibinfo{person}{Jermey~B. Merrill}.} \bibinfo{year}{2023}\natexlab{}.
\newblock \showarticletitle{17 fatalities, 736 crashes: The shocking toll of Tesla’s Autopilot}.
\newblock \bibinfo{howpublished}{\url{https://www.washingtonpost.com/technology/2023/06/10/tesla-autopilot-crashes-elon-musk/}}.
\newblock  (\bibinfo{year}{2023}).
\newblock


\bibitem[Spiers(2024)]%
        {nyt-gpt-reminds}
\bibfield{author}{\bibinfo{person}{Elizabeth Spiers}.} \bibinfo{year}{2024}\natexlab{}.
\newblock \showarticletitle{I Finally Figured Out Who ChatGPT Reminds Me Of}.
\newblock \bibinfo{howpublished}{\url{https://www.nytimes.com/2024/01/07/opinion/chatgpt-generative-ai.html}}.
\newblock  (\bibinfo{year}{2024}).
\newblock


\bibitem[Thomas(2006)]%
        {thomas2006general}
\bibfield{author}{\bibinfo{person}{David~R Thomas}.} \bibinfo{year}{2006}\natexlab{}.
\newblock \showarticletitle{A general inductive approach for analyzing qualitative evaluation data}.
\newblock \bibinfo{journal}{\emph{American journal of evaluation}} \bibinfo{volume}{27}, \bibinfo{number}{2} (\bibinfo{year}{2006}), \bibinfo{pages}{237--246}.
\newblock


\bibitem[Waytz et~al\mbox{.}(2010)]%
        {waytz2010sees}
\bibfield{author}{\bibinfo{person}{Adam Waytz}, \bibinfo{person}{John Cacioppo}, {and} \bibinfo{person}{Nicholas Epley}.} \bibinfo{year}{2010}\natexlab{}.
\newblock \showarticletitle{Who sees human? The stability and importance of individual differences in anthropomorphism}.
\newblock \bibinfo{journal}{\emph{Perspectives on Psychological Science}} \bibinfo{volume}{5}, \bibinfo{number}{3} (\bibinfo{year}{2010}), \bibinfo{pages}{219--232}.
\newblock


\bibitem[Wei et~al\mbox{.}(2022)]%
        {wei2022chain}
\bibfield{author}{\bibinfo{person}{Jason Wei}, \bibinfo{person}{Xuezhi Wang}, \bibinfo{person}{Dale Schuurmans}, \bibinfo{person}{Maarten Bosma}, \bibinfo{person}{Fei Xia}, \bibinfo{person}{Ed Chi}, \bibinfo{person}{Quoc~V Le}, \bibinfo{person}{Denny Zhou}, {et~al\mbox{.}}} \bibinfo{year}{2022}\natexlab{}.
\newblock \showarticletitle{Chain-of-thought prompting elicits reasoning in large language models}.
\newblock \bibinfo{journal}{\emph{Advances in Neural Information Processing Systems}}  \bibinfo{volume}{35} (\bibinfo{year}{2022}), \bibinfo{pages}{24824--24837}.
\newblock


\bibitem[Xi et~al\mbox{.}(2023)]%
        {xi2023rise}
\bibfield{author}{\bibinfo{person}{Zhiheng Xi}, \bibinfo{person}{Wenxiang Chen}, \bibinfo{person}{Xin Guo}, \bibinfo{person}{Wei He}, \bibinfo{person}{Yiwen Ding}, \bibinfo{person}{Boyang Hong}, \bibinfo{person}{Ming Zhang}, \bibinfo{person}{Junzhe Wang}, \bibinfo{person}{Senjie Jin}, \bibinfo{person}{Enyu Zhou}, {et~al\mbox{.}}} \bibinfo{year}{2023}\natexlab{}.
\newblock \showarticletitle{The rise and potential of large language model based agents: A survey}.
\newblock \bibinfo{journal}{\emph{arXiv preprint arXiv:2309.07864}} (\bibinfo{year}{2023}).
\newblock


\bibitem[Xiang(2023)]%
        {vice_zuckerberg}
\bibfield{author}{\bibinfo{person}{Chloe Xiang}.} \bibinfo{year}{2023}\natexlab{}.
\newblock \bibinfo{title}{Zuckerberg's Vision for AI: A Bot That Makes Ads and Helps You Say Happy Birthday to Your Friends}.
\newblock
\newblock
\urldef\tempurl%
\url{https://www.vice.com/en/article/jg55j7/zuckerbergs-vision-for-ai-a-bot-that-makes-ads-and-helps-you-say-happy-birthday-to-your-friends}
\showURL{%
\tempurl}
\newblock
\shownote{Accessed on: 9 October 2023}.


\bibitem[Yagoda and Gillan(2012)]%
        {yagoda2012you}
\bibfield{author}{\bibinfo{person}{Rosemarie~E Yagoda} {and} \bibinfo{person}{Douglas~J Gillan}.} \bibinfo{year}{2012}\natexlab{}.
\newblock \showarticletitle{You want me to trust a ROBOT? The development of a human--robot interaction trust scale}.
\newblock \bibinfo{journal}{\emph{International Journal of Social Robotics}}  \bibinfo{volume}{4} (\bibinfo{year}{2012}), \bibinfo{pages}{235--248}.
\newblock


\bibitem[Zhang et~al\mbox{.}(2020)]%
        {zhang2020effect}
\bibfield{author}{\bibinfo{person}{Yunfeng Zhang}, \bibinfo{person}{Q~Vera Liao}, {and} \bibinfo{person}{Rachel~KE Bellamy}.} \bibinfo{year}{2020}\natexlab{}.
\newblock \showarticletitle{Effect of confidence and explanation on accuracy and trust calibration in AI-assisted decision making}. In \bibinfo{booktitle}{\emph{Proceedings of the 2020 conference on fairness, accountability, and transparency}}. \bibinfo{pages}{295--305}.
\newblock


\end{thebibliography}

\appendix
\section{IDAQ Score Across Participants}
\label{app:idaq-participants}
\begin{figure}[h] 
    \centering
    \includegraphics[width=0.5\textwidth]{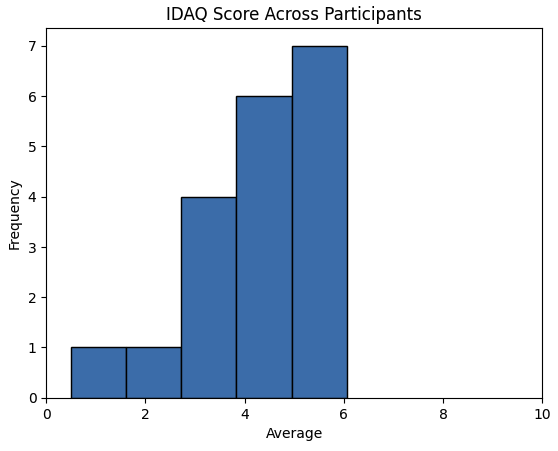}
    \caption{IDAQ score distribution across participants (score range: 0-10, with 10 corresponding to greatest tendency to anthropomorphize).}
    \label{fig:idaq}
\end{figure}

\section{Additional observations} \label{app:qualitative-observations}

\subsection{Challenges of Judging reliability, trustworthiness, or consistency with one Response:}
\label{sec:unabletojudge}

Six out of 19 participants indicated that being presented with a single response from the model was insufficient for assessing its reliability, trustworthiness, or consistency. For example, Participant 4 expressed uncertainty when presented with only one answer, stating that they would need to ask the model multiple questions to gain a more comprehensive understanding. Participant 16 also voiced skepticism when presented with just one response, hinting at a broader trend of user caution in these situations. On a related note, Participant 7 demonstrated how perceptions about the system could evolve with exposure to more responses. Initially, Participant 7 placed a high level of trust in the single response generated by the model, largely because it was the only data point available to them at that moment. They stated, "When I saw the single sample, I put more credence into it because it was the only thing I had." However, as they encountered additional responses, they recognized the limitations of relying solely on one answer, stating: "But then, the more I realized that this one sample is useless, and the 10 samples are actually giving me a pretty good idea of what it thinks." The multiple samples offered a more comprehensive view of the model's capabilities and reasoning, leading Participant 7 to revise their initial assessment.

\subsection{Seeing more responses increased some participants' perceptions of their understanding of the model's reasoning process and appreciation of its knowledge}
Our study found that being presented with multiple system responses enhances users' perception of their understanding of the system's reasoning process and appreciation of its knowledge. For example, Participant 6 observed the consistent format/schema across multiple responses and believed it made it easier for them to grasp the model's operational logic---the set of mechanisms governing how the model arrives at a particular answer or output---thereby boosting their confidence in the system's reasoning process. They stated, "When I was looking at more responses, it became easier for me to understand how the model is working. There was some level of consistency in the format of the responses, which made me a bit more confident in the model's reasoning process." Participants 16 and 17 also expressed a positive view on the impact of multiple responses, noting that a range of answers effectively showcased the system's expansive knowledge base. "Especially when there are more responses, we can see more knowledge from that part," stated Participant 16. Similarly, Participant 17 commended the system's ability to offer rephrased and less commonly known facts, describing it as "really knowledgeable." Further extending this discussion, Participant 19 stated that, "the various ways a sentence can be structured give me more of an idea of its decision-making processes."

\subsection{Factors that contributed to machine-like perceptions}
Participants noted that some of the generated responses were machine-like because a human would not respond with similar statements. Participant 4 said “I think it's pretty machine-like because it doesn't give us much. It doesn't give credit to other people or explain what they did. While the main person might have the biggest contribution, I think people are generally nicer and would attribute the success to others.” Others deemed responses to be machine-like because of the style and structure. Participant 11 noted “it's a bit too formal, it gives me a machine-like feeling as well. It's like when I ask an expert or a professor; they would probably answer me more colloquially. The machine's responses are like very well-written answers in a textbook, so it feels a bit machine-like in that way.” Participant 17 shared a similar sentiment, “I don't think they're human-like, so the generations are still kind of like standard, like very similar to what say, retrieval system gives you or how Wikipedia will write this knowledge.”

When considering responses to the advice based question, participants thought that responses were machine-like because they were too concise and general (could apply to anyone asking the question). In addition, participants noted that a human would incorporate the context provided in the query when answering the question. Participant 13 captured this by stating “ it just felt like it was like spitting out something that was too vague and concise for it to be like, it made me think that again, like it's not balancing concision with our conciseness with, like, being specific on what I'm asking.” Participant 15 noted that the advice provided “seemed very impersonal.” In contrast, for the factual based question (without justification), participant noted that long responses with more detail made the responses more machine-like; participant 9 noted “...longer responses make it seem like it's looking up something instead of like really conversing with you because in conversations that we ask this question, we do not talking as much.”  Similarly, participant1 noted, “... a human would simply say Thomas Edison.” Other participants mentioned that the responses were machine-like because they did not express empathy and emotion or ask follow-up questions. 

\section{Standardized Survey Questions}

\subsection{Need for Cognition Scale (NCS-6)}
The following questions are on a 7 point likert scale: 
\label{app:ncs-6}
\begin{enumerate}
    \item I would prefer complex to simple problems. 
    \item I like to have the responsibility of handling a situation that requires a lot of thinking.
    \item Thinking is not my idea of fun.
    \item I would rather do something that requires little thought than something that is sure to challenge my thinking abilities.
    \item I really enjoy a task that involves coming up with new solutions to problems. 
    \item I would prefer a task that is intellectual, difficult, and important to one that is somewhat important but does not require much thought.
\end{enumerate}

\subsection{The Individual Differences in Anthropomorphism Questionnaire (IDAQ)}
The following questions are on an 11 point likert scale:
\label{app:idaq}
\begin{enumerate}
    \item To what extent does the average robot have consciousness?
    \item To what extent does the average fish have free will?
    \item To what extent does the average mountain have free will?
    \item To what extent does a television set experience emotions?
    \item To what extent does the average robot have consciousness?
    \item To what extent do cows have intentions?
    \item To what extent does a car have free will?
    \item To what extent does the ocean have consciousness?
    \item To what extent does the average computer have a mind of its own?
    \item To what extent does a cheetah experience emotions?
    \item To what extent does the environment experience emotions?
    \item To what extent does the average insect have a mind of its own?
    \item To what extent does a tree have a mind of its own?
    \item To what extent does the wind have intentions?
    \item To what extent does the average reptile have consciousness?
\end{enumerate}

\subsection{Godspeed Questionnaire}
The following questions are on a 5 point likert scale:
\label{app:godspeed}
\begin{enumerate}
    \item How would you rate the responses from "Fake" to "Natural"?
    \item How would you rate the responses from "Machinelike" to "Humanlike"?
    \item How would you rate the responses from "Unconscious" to "Conscious"?
    \item How would you rate responses from "Artificial" to "Lifelike"?
\end{enumerate}

\section{Human Computer Trust (HCT) Scale}
The following questions are on a 7 point likert scale: \\
\label{app:hct}
Perceived Reliability
\begin{enumerate} 
    \item The system always provides the advice I require to make my decision.	
    \item The system performs reliably	
    \item The system responds the same way under the same conditions at different times.	
    \item I can rely on the system to function properly.	
    \item The system analyzes problems consistently.	
\end{enumerate}

Perceived Technical Competence
\begin{enumerate}
    \item The system uses appropriate methods to reach decisions.	
    \item The system has sound knowledge about this type of problem built into it.	
    \item The advice the system produces is as good as that which a highly competent person could produce.	
    \item The system correctly uses the information I enter.	
    \item The system makes use of all the knowledge and information available to it to produce its solution to the problem.
\end{enumerate}
        	
Perceived Understandability
\begin{enumerate}
    \item I know what will happen the next time I use the system because I understand how it behaves.	
    \item I understand how the system will assist me with decisions I have to make. 	
    \item Although I may not know exactly how the system works, I know how to use it to make decisions about the problem.	\item It is easy to follow what the system does.	
    \item I recognize what I should do to get the advice I need from the system the next time I use it. 
\end{enumerate}
         
Faith
\begin{enumerate}
    \item I believe advice from the system even when I don’t know for certain that it is correct.	
    \item When I am uncertain about a decision I believe the system rather than myself.	
    \item If I am not sure about a decision, I have faith that the system will provide the best solution. 	
    \item When the system gives unusual advice I am confident that the advice is correct.	
    \item Even if I have no reason to expect the system will be able to solve a difficult problem, I still feel certain that it will. 
\end{enumerate}
    	 
Personal Attachment
\begin{enumerate}
    \item I would feel a sense of loss if the system was unavailable and I could no longer use it. 	 
    \item I feel a sense of attachment to using the system. 	
    \item I find the system suitable to my style of decision-making. 	
    \item I like using the system for decision-making. 	 
    \item I have a personal preference for making decisions with the system. 
\end{enumerate}

\section{Semi-structured Interview Questions}
The following questions were asked after the 3 conditions were completed:
\label{app:interview}

General Experience
\begin{enumerate}
    \item Was there anything about the generated response(s) that surprised you?
    \item Can you summarize your overall experience interacting with the language model?
    \item For each condition, can you summarize your experience interacting with the language model?
    \item Did you notice any differences in your experience across the conditions?
\end{enumerate}

Anthropomorphism
\begin{enumerate}
    \item Did you find yourself attributing human-like qualities to the language model? Can you give examples?
    \item Did you notice any variations in your attributions of human-like qualities across the conditions? 
\end{enumerate}
     
Trust and Credibility
\begin{enumerate}
    \item How much did you trust the information and suggestions provided by the language model?
    \item How did you feel about the trustworthiness of the information and suggestions provided by the language model in each condition?
    \item Were there specific moments or conditions that increased or decreased your trust? 
\end{enumerate}
        
Perceived Intelligence:
\begin{enumerate}
     \item Did you find the language model to be knowledgeable and competent? Why or why not?
    \item Were there noticeable differences in your perception of the language model's intelligence across conditions? 
\end{enumerate}
       
Future Interaction:
\begin{enumerate}
    \item Would you use this language model again for similar tasks? Why or why not?
    \item Did any particular condition influence your willingness to use the language model in the future? 
\end{enumerate}
  
Open-Ended: 
\begin{enumerate}
    \item Is there anything else you'd like to share about how your experience or perceptions differed across the conditions?
    \item What changes or improvements would you suggest for a better user experience?
\end{enumerate}

\section{Quantitative Results}
\label{app:quantitiave-results}
The following graphs show the difference in participant responses to pre-study survey questions, evaluated based on the question type assigned.

\begin{figure}[h] 
    \centering
    \includegraphics[width=0.5\textwidth]{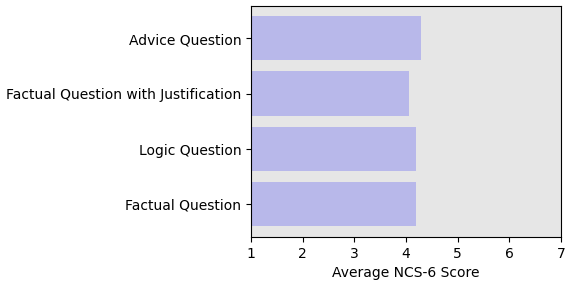}
    \caption{Average Need for Cognition grouped by question type}
    \label{fig:ncs-acrossq}
\end{figure}

\begin{figure}[h] 
    \centering
    \includegraphics[width=0.5\textwidth]{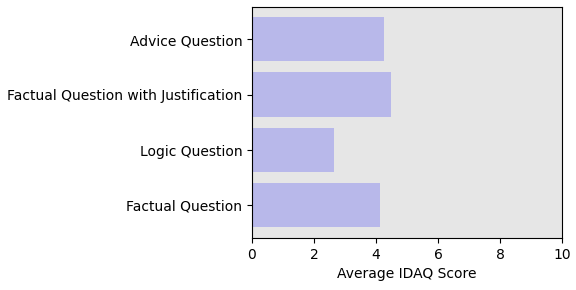}
    \caption{IDAQ score distribution across question type (score range: 0-10, with 10 corresponding to greatest tendency to anthropomorphize).}
    \label{fig:idaq-acrossq}
\end{figure}

\begin{figure}[h] 
    \centering
    \includegraphics[width=0.5\textwidth]{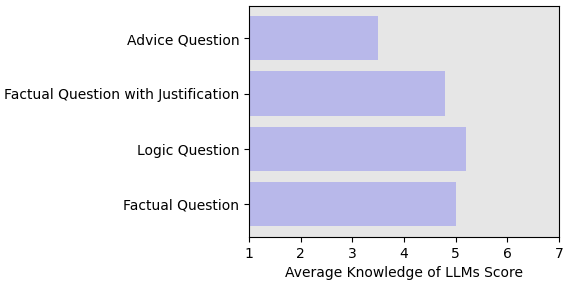}
    \caption{Participants were asked: "How much do you know about large language models (LLMs) or systems powered by such models?" This graph reflects the average score grouped by question type.}
    \label{fig:knowledge-acrossq}
\end{figure}

\begin{figure}[h] 
    \centering
    \includegraphics[width=0.5\textwidth]{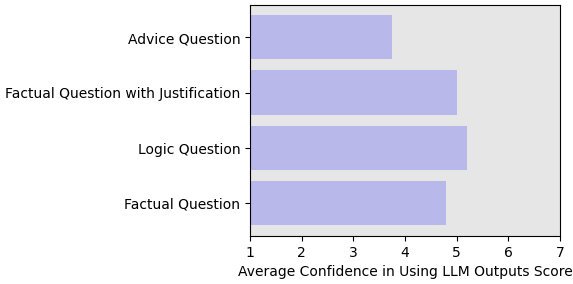}
    \caption{Participants were asked: "How confident are you in using the outputs from a LLM?" This graph reflects the average score grouped by question type.}
    \label{fig:confidence-acrossq}
\end{figure}

\begin{figure}[h] 
    \centering
    \includegraphics[width=0.5\textwidth]{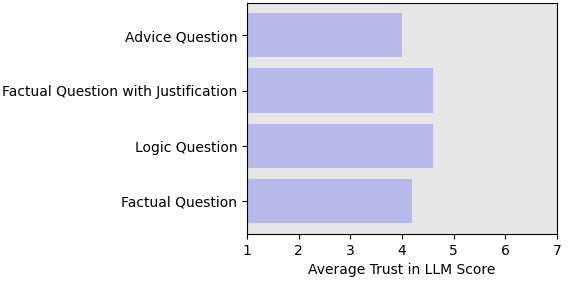}
    \caption{Participants were asked: "How much do you trust LLMs to perform a task accurately or provide accurate information?" This graph reflects the average score grouped by question type.}
    \label{fig:trust-acrossq}
\end{figure}

\end{document}